\documentstyle[aps]{revtex}
\begin{document}
\preprint{UTF 340 - Dec. 1994}
\draft
\title{
Wannier functions and Fractional Quantum Hall Effect}
\author{
R. Ferrari}
\address{Dipartimento di Fisica
Universit\'a di Trento, 38050 Povo, Trento Italy \\
and \\I.N.F.N. sezione
di
Padova, gruppo collegato di Trento\\
rferrari@alpha.science.unitn.it}
\date{\today}
\maketitle
\begin{abstract} We introduce and study the Wannier functions for
an electron moving in a plane under the influence of a perpendicular
uniform and constant magnetic field. The relevance for
the Fractional Quantum Hall Effect is discussed; in particular
it shown that an interesting Hartree-Fock state can be constructed
in terms of Wannier functions.
\end{abstract}
\rightline{cond-mat/9501055}
\widetext
\section{Introduction}
It is widely accepted that Quantum Hall Effect \cite{klitzing,tsui}
can be described by a model of interacting electrons moving in a plane.
For many features of the phenomenon the zero temperature limit
seems to be a good approximation.
\par
In the attempt to describe the state in terms of single
particle approximation and subsequent perturbations around
this trial state, the author has introduced \cite{ferrari1}
a set of functions which are eigenfunctions of the single particle
Hamiltonian and moreover are invariant, up to a phase,
under finite translations associated to a lattice.
The phase depends on the coordinates. In subsequent papers
\cite{ferrari2,ferrari3} an {\sl Ansatz} was introduced in
order to
construct the trial state. In Ref. \cite{ferrari3} numerical evidence
is given that this state yields a mean Coulomb energy close
to the value provided by genuine many-body states (e.g. Laughlin's
\cite{laughlin}).
{}From this property and from the fact that the trial state is given
in analytic form in terms of a Slater determinant
(or single particle approximation)
the author hopes that a perturbative approach based on this
{\sl Ansatz} will improve the value of the mean energy and
provide a further progress in the  understanding of the phenomenon.
\par
In this note the Wannier functions are introduced by using the
functions invariant under translations. This construction has
some degree of arbitrariness related to the phase, which cannot
be fixed {\sl a priori}. Some possible natural choices are considered
and numerical methods are used in order to decide which is the
better one.
\par
Finally the Wannier functions are used in order to reformulate
the {\sl Ansatz} for the trial state. This analysis shows that
there is at least another natural candidate for the trial state.
By considering the particular case of filling factor $1/3$ numerical
evidence is given that the original proposal gives the better choice.
\section{The formalism}
\label{sec:formalism}
In this section some essential points developed in I and III
are briefly reviewed.
The Hamiltonian for the single electron is
\begin{equation}
H = {1\over 2} \left[
\left( -i\partial_x -{y\over 2}\right )^2 +
\left( -i\partial_y +{x\over 2}\right )^2
\right ]
\label{1}
\end{equation}
where the unit of length is
\begin{equation}
\lambda =\left({{c\hbar}\over {eB}}\right)^{1\over 2}
\label{2}
\end{equation}
and the unit of energy
\begin{equation}
\hbar\omega_{\rm c} = \hbar{{eB}\over {mc}}.
\label{3}
\end{equation}
The complex notation is used for the vectors
\begin{equation}
w = w_1+iw_2.
\label{4}
\end{equation}
Step down and step operators are defined
\begin{eqnarray}
&&a\equiv {1\over{\sqrt{2}}}\left(
{{w^\ast}\over{2}}+ 2\partial_w
\right)
\nonumber \\
&&b\equiv {1\over{\sqrt{2}}}\left(
{{w}\over{2}}+ 2\partial_{w^\ast}
\right).
\label{6}
\end{eqnarray}
They obey the algebra
\begin{eqnarray}
&&[a,a^\dagger] = 1 \qquad [b,b^\dagger] = 1
\nonumber \\
&&[a,b^\dagger] = 0 \qquad [a,b] = 0
\label{8}
\end{eqnarray}
and moreover
\begin{equation}
H = a^\dagger a + {1\over 2}.
\label{9}
\end{equation}
Let $\varphi_{n_L}$ be
the solution of
\begin{equation}
a^\dagger a \varphi_{n_L} = n_L \varphi_{n_L} \qquad
b\varphi_{n_L} = 0.
\label{10}
\end{equation}
One gets easily
\begin{equation}
\varphi_{n_L}({\bf r}) = (2\pi n_L!)^{-{1\over 2}}
\left( {{x+iy}\over{\sqrt{2}}}
\right)^{n_L}
\exp\left(-{{r^2}\over{4}}\right).
\label{11}
\end{equation}
It is convenient to introduce the coherent-state operator
\begin{equation}
S(w) = \exp\left({{1}\over{\sqrt{2}}}
(w^\ast b-w b^\dagger)\right)=
\exp\left({i\over 2}{\bf w}\times{\bf r}\cdot
{\hat{\bf z}}\right )
\exp\left (w_1\partial_x + w_2\partial_y \right )
\label{12}
\end{equation}
i.e. it is the product of a translation and a phase. It has the
properties
\begin{equation}
[S(w),H] = 0
\label{13}
\end{equation}
and
\begin{equation}
S(c)S(d) = S(c+d) \exp\left(-{i\over 2}{\bf c}\times{\bf d}\cdot
{\hat{\bf z}}\right )
\label{14}
\end{equation}
(Magnetic Translation Group \cite{zak} (MTG)).
Thus they commute
\begin{equation}
[S(c),S(d)] = 0\qquad {\rm if}~~
{\bf c}\times{\bf d}\cdot
{\hat{\bf z}}=2\pi u \qquad(u~~{\rm integer}).
\label{15}
\end{equation}
Then these operators are adequate in order to impose
quasi-periodic
boundary conditions on a parallelogram $A$ with side vectors
${\bf L}_1,{\bf L}_2$
\begin{eqnarray}
&&S(L_1)\psi = e^{i\theta_1}\psi
\nonumber \\
&&S(L_2)\psi = e^{i\theta_2}\psi
\label{16}
\end{eqnarray}
provided they commute
\begin{equation}
[S(L_1),S(L_2)] = 0\qquad {\rm i.e.}~~
{\bf L}_1\times{\bf L}_2\cdot
{\hat{\bf z}}=2\pi g_L \qquad(g_L~~{\rm integer}).
\label{17}
\end{equation}
Then $g_L$ will be the degeneracy of the Landau levels. The elements of the
MTG are defined by the constraint
\begin{equation}
[S(w),S(L_j)] = 0\qquad  j = 1,2
\label{17.1}
\end{equation}
i.e.
\begin{equation}
w = {{1}\over{g_L}}(-n_1 L_2 + n_2 L_1) \qquad {\rm with}
\quad n_1,n_2 ~~{\rm integers}.
\label{17.2}
\end{equation}
\par
Let $c,d$ be any two vectors that satisfy eq. (\ref{15}) and
provide a commensurate tiling of the parallelogram, i.e.
\begin{eqnarray}
&&L_1 = p c
\nonumber \\
&&L_2 = p' c + q d
\label{18}
\end{eqnarray}
where the coefficients are integers. One has
\begin{equation}
g_L = pq u.
\label{18.1}
\end{equation}
A finer lattice can be introduced generated by the
vectors $\bf f$ and $\bf g$ such that
\begin{eqnarray}
{\bf c}& =& \kappa{\bf f} + \iota '{\bf g} \nonumber\\
{\bf d}& =& \kappa '{\bf f} +\iota {\bf g}
\label{19}
\end{eqnarray}
with the condition that the coefficients are integer numbers and
moreover
\begin{equation}
{\bf f}\times {\bf g}\cdot {\hat{\bf z}} = 2\pi
\label{20}
\end{equation}
i.e.
\begin{equation}
\kappa\iota - \kappa '\iota ' = u.
\label{21}
\end{equation}
\par
Sets of functions can be introduced which are eigenvectors of the
Hamiltonian and are invariant under translations given by
the sites of the lattice generated by
$\bf c, d$ or by $\bf f, g$. Thus solutions of
\begin{eqnarray}
&&S(c)\psi_{n_L}^{\mu\nu} = e^{i\mu}\psi_{n_L}^{\mu\nu}
\nonumber \\
&&S(d)\psi_{n_L}^{\mu\nu} = e^{i\nu}\psi_{n_L}^{\mu\nu}
\label{22}
\end{eqnarray}
and
\begin{eqnarray}
&&S(f)\phi_{n_L}^{\alpha\beta} = e^{i\alpha}\phi_{n_L}^{\alpha\beta}
\nonumber \\
&&S(g)\phi_{n_L}^{\alpha\beta} = e^{i\beta}\phi_{n_L}^
{\alpha\beta}
\label{22.1}
\end{eqnarray}
can be introduced.
The solutions of the above equations are given by
\begin{eqnarray}
\psi^{\mu\nu}_{n_L} ( {\bf r} ) &=
&( p q )^{-{1\over{2}}}\sum_{m,n=-\infty}^{+\infty}
[ S (c) e^{- i \mu} ]^m
[ S (d) e^{- i \nu} ]^n ~{\varphi}_{n_L} ({\bf r} )
\nonumber\\
&
=&(pq)^{-{1\over{2}}} \sum_{m,n=-\infty}^{+\infty} ( - )^{mnu}
\exp [ - i (\mu m + \nu n ) ]
\nonumber\\
&&\exp \Bigl[ {i\over{2}} {\bf{\hat z}} \cdot
(m {\bf c} + n {\bf d} ) \times {\bf r} \Bigl]
{}~{\varphi}_{n_L} ( {\bf r} + m {\bf c} + n
{\bf d} )
\label{23}
\end{eqnarray}
and similarly
\begin{eqnarray}
\phi_{n_L}^{\alpha\beta} &= &
( g_L)^{-{1\over{2}}} \sum_{m,n=-\infty}^{+\infty}
[ S (f) e^{- i \alpha} ]^m
[ S (g) e^{- i \beta} ]^n ~{\varphi}_{n_L} ({\bf r} )
\nonumber\\
&=&(g_L)^{-{1\over{2}}} \sum_{m,n=-\infty}^{+\infty} ( - )^{mn}
\exp [ - i (m\alpha+ n\beta ) ]
\nonumber\\
&&\exp \Bigl[ {i\over{2}} {\bf{\hat z}} \cdot
(m {\bf f} + n {\bf g} ) \times {\bf r} \Bigl]
{}~{\varphi}_{n_L} ( {\bf r} + m {\bf f} + n
{\bf g} ).
\label{24}
\end{eqnarray}
The value of the parameters are fixed by the boundary conditions
in eq. (\ref{16}). See eqs. (49), (83) and (84) of I.
In particular it is easy to find the following relations as a consistency
condition
for the eqs. (\ref{22}) and (\ref{22.1})
\begin{eqnarray}
2\pi n_1 + \mu&= & \pi\kappa\iota'+\kappa\alpha
+\iota'\beta
\nonumber\\
2\pi n_2 + \nu&= & \pi\kappa'\iota+\kappa'\alpha
+\iota\beta.
\label{24.1}
\end{eqnarray}
The norm of the wave function is (see eqs. (69) and (91) of I)
\begin{equation}
\Vert\phi_{n_L}^{\alpha\beta}\Vert^2 =
\sum_{m,n=-\infty}^{+\infty} \int_{{\cal R}^2} d^2 r \{
[ S (f)]^m
[ S (g)]^n ~{\varphi}_{n_L} ({\bf r} )\}^\ast
{\varphi}_{n_L} ({\bf r} ) e^{i (m\alpha + n\beta)} .
\label{25}
\end{equation}
One gets
\begin{equation}
\Vert\phi_{n_L}^{\alpha\beta}\Vert^2 =
\sum_{m,n=-\infty}^{+\infty}  ( - )^{mn}e^{i (m\alpha + n\beta)}
\exp(-{1\over 4}|m f + n g|^2).
\label{26}
\end{equation}
A similar result is valid for the other set of functions
\begin{equation}
\Vert\psi_{n_L}^{\mu\nu}\Vert^2 =
\sum_{m,n=-\infty}^{+\infty}  ( - )^{mnu}e^{i (m\mu + n\nu)}
\exp(-{1\over 4}|m c + n d|^2).
\label{26.1}
\end{equation}
The set of functions $\{\phi_{n_L}^{\alpha\beta}\}$ form a
complete set.
We use capital letters to denote the normalized functions
\begin{eqnarray}
\Phi_{n_L}^{\alpha\beta} &\equiv &
{{\phi_{n_L}^{\alpha\beta}}
\over{\Vert\phi_{n_L}^{\alpha\beta}\Vert}}
\nonumber\\
\Psi_{n_L}^{\mu\nu} &\equiv &
{{\psi_{n_L}^{\mu\nu}}
\over{\Vert\psi_{n_L}^{\mu\nu}\Vert}}.
\label{27}
\end{eqnarray}
\par
For computation of the matrix elements of the Coulomb part
(in II and  III) it was convenient to introduce a new set
of functions  $\{{\hat\Phi}_{n_L}^{\alpha\beta}\}$. One considers
a reference wave function  $\Phi_{n_L}^{\alpha_0\beta_0}$ and
defines (see Appendix in III)
\begin{equation}
{\hat\Phi}_{n_L}^{\alpha\beta} \equiv
S(w_{\alpha\beta})\Phi_{n_L}^{\alpha_0\beta_0}
\label{28}
\end{equation}
where $w_{\alpha\beta}$ is a standard set of vectors
\begin{equation}
w_{\alpha\beta}= (2\pi)^{-1}\left[\left(\beta-\beta_0\right)f -
\left(\alpha -\alpha_0\right)g\right].
\label{29}
\end{equation}
By using the composition rules in eq. (\ref{14})
\begin{eqnarray}
S(w_{\alpha\beta})S(f) &=&S(f)S(w_{\alpha\beta})
\exp\left(-i{\bf w}_{\alpha\beta}\times{\bf f}\cdot
{\hat{\bf z}}\right )=S(f)S(w_{\alpha\beta})
\exp(-i(\alpha-\alpha_0))
\nonumber\\
S(w_{\alpha\beta})S(g) &=&S(g)S(w_{\alpha\beta})
\exp\left(-i{\bf w}_{\alpha\beta}\times{\bf g}\cdot
{\hat{\bf z}}\right )=S(g)S(w_{\alpha\beta})
\exp(-i(\beta-\beta_0))
\nonumber\\
\label{30}
\end{eqnarray}
it is easy to show that ${\hat\Phi}_{n_L}^{\alpha\beta}$
satisfies eqs. (\ref{22.1}) and therefore differs from
$\Phi_{n_L}^{\alpha\beta}$ by a phase.
\section{Wannier functions}
\label{sec:wannier}
In this section Wannier functions
\cite{wa1,k1,wa2,bl,cl1,cl2,cl3,nen,gel}
are introduced for a particle
moving in a constant and homogeneous magnetic field.
Various sets of Wannier functions can be introduced.
By starting from
$\{\Phi_{n_L}^{\alpha\beta}\}$ one can define (the index
${n_L}$ will be suppressed for convenience of notation)
\begin{equation}
{\cal W}_{r,s}\equiv g_L^{-{1\over 2}}\sum_{\alpha\beta}
e^{-i (r\alpha + s\beta)}\Phi^{\alpha\beta}.
\label{31}
\end{equation}
If one uses the set $\{{\hat\Phi}_{n_L}^{\alpha\beta}\}$ a different
Wannier functions is obtained
\begin{equation}
{\hat{\cal W}}_{r,s}\equiv g_L^{-{1\over 2}}\sum_{\alpha\beta}
e^{-i (r\alpha + s\beta)}{\hat\Phi}^{\alpha\beta}.
\label{32}
\end{equation}
Yet another Wannier function is obtained if
 ${\hat\Phi}_{n_L}^{\alpha\beta}$ is replaced by
\begin{equation}
S(w_{\alpha_0\beta})S(w_{\alpha\beta_0})
\Phi_{n_L}^{\alpha_0\beta_0}=
\exp({i\over{4\pi}}(\alpha-\alpha_0)(\beta-\beta_0))
{\hat\Phi}_{n_L}^{\alpha\beta}.
\label{33}
\end{equation}
All these definitions provide a set of orthonormal functions
located around the lattice site
\begin{equation}
{\bf x} = r{\bf f} + s{\bf g}.
\label{34}
\end{equation}
Moreover they have the property (see eq. (\ref{22.1}))
\begin{eqnarray}
S(r'f+s'g){\cal W}_{r,s}({\bf x})&=&
\exp({i\over 2}(r'{\bf f}+s'{\bf g})\times {\bf x}\cdot
{\hat{\bf z}}){\cal W}_{r,s}({\bf x}+r'{\bf f}+s'{\bf g})
\nonumber \\
&=& (-)^{r's'}{\cal W}_{(r-r'),(s-s')}({\bf x}).
\label{35}
\end{eqnarray}
\par
One expects that any different choice of phases for the
translation
invariant functions provides Wannier functions with different
shapes \cite{k1}. By numerical inspection one can chose
the functions with
better localization properties among the three proposal given in eqs.
(\ref{31}), (\ref{32}) and (\ref{33}). From Figs. \ref{w1},
\ref{w2} and \ref{w3} one sees clearly that
the choice in eq. (\ref{31}) (${\cal W}$) gives the best
localization.
\par
The formalism developed in section \ref{sec:formalism} allows the
introduction of other kind of Wannier functions, i.e. those
associated to the lattice generated by the $c,d$ vectors
\cite{evarestov}.
\begin{equation}
{\cal W}_{m,n}^{(u)}\equiv \left({{u}
\over{g_L}}\right)^{{1\over 2}}\sum_{\mu\nu}
e^{-i (m\mu+n\nu)}\Psi^{\mu\nu}
\label{36}
\end{equation}
is expected to be localized around the lattice site
\begin{equation}
{\bf x} = m{\bf c} + n{\bf d}.
\label{37}
\end{equation}
Fig. \ref{w4} shows the shape of the Wannier function
${\cal W}_{m,n}^{(u)}$. It should be noticed that the functions
$\{{\cal W}_{m,n}^{(u)}\}$ do not form a complete set of functions.
\par
It is interesting to elaborate further the difference between the
sets $\{{\cal W}\}$ and $\{{\cal W}^{(u)}\}$. It can be shown
(see eq. (90) of I) that
\begin{equation}
\psi_{n_L}^{\mu\nu} = u^{-{1\over{2}}}
\sum_{\alpha\beta}^{{\rm fixed}~\mu\nu}
\phi_{n_L}^{\alpha\beta}.
\label{38}
\end{equation}
On the other side, $\{{\cal W}_{r,s}\}$ evaluated on a $c,d$ lattice
site
\begin{eqnarray}
r &=& m\kappa  + n\kappa'
\nonumber \\
s&=& m\iota' + n\iota,
\label{39}
\end{eqnarray}
gives (see eq. (\ref{24.1}))
\begin{eqnarray}
&&r\alpha + s\beta = m(\kappa\alpha+\iota'\beta)
+ n(\kappa'\alpha + \iota\beta)
\nonumber \\
&&= m[\mu-\pi\kappa\iota'+2\pi n_1] +
n[\nu-\pi\kappa'\iota + 2\pi n_2]
\label{40}
\end{eqnarray}
and therefore
\begin{equation}
{\cal W}_{r,s}=\exp(-i\pi(m\kappa\iota'+n\kappa'\iota))
\left({{u}\over{g_L}}\right)^{{1\over 2}}\sum_{\mu\nu}
e^{-i (m\mu+n\nu)} \left\{u^{-{1\over{2}}}
\sum_{\alpha\beta}^{{\rm fixed}~\mu\nu}
\Phi_{n_L}^{\alpha\beta}\right\}
\label{41}
\end{equation}
on the sites of the $c,d$ lattice.
\par
The conclusion of this section is that one has two good
candidates ($\{{\cal W}_{r,s}\}$ and
$\{{\cal W}_{m,n}^{(u)}\}$) which can be used for the
construction of a trial state for the many-body problem.
\section{Trial state for FQHE}
\label{sec:fqhe}
The original {\sl Ansatz} for the FQHE trial state presented
in the papers II and III was formulated in terms of the
wave function $\Psi^{\mu\nu}$ of eqs. (\ref{23}) and (\ref{27}).
For the filling factor $1/u$ it is just the Slater determinant
of the $g_L/u$ functions with $n_L=0$. In the second quantization
formalism where $\psi$ is fermion field (spin is neglected),
the trial state is
\begin{equation}
|\Omega\rangle = \prod_{\mu\nu}\int_{A}d^2x \psi^\dagger({\bf x})
\Psi^{\mu\nu}({\bf x})|0\rangle.
\label{42}
\end{equation}
The equation (\ref{36}) is a unitary transformation, thus,
up to a phase, the state $|\Omega\rangle$ can be written in terms
of Wannier functions
\begin{equation}
|\Omega\rangle \simeq \prod_{m,n}\int_{A}
d^2x \psi^\dagger({\bf x})
{\cal W}_{m,n}^{(u)}({\bf x})|0\rangle
\label{43}
\end{equation}
where $m,n$ are the sites of the lattice generated by $c,d$.
\par
{}From the discussion at the end of section \ref{sec:wannier} it
follows that one can similarly construct a trial state in terms
of the Wannier functions ${\cal W}_{r,s}$ given in eq. (\ref{31})
\begin{equation}
|{\tilde\Omega}\rangle \equiv \prod_{m,n}\int_{A}
d^2x \psi^\dagger({\bf x})
{\cal W}_{r,s}({\bf x})|0\rangle\Big |_
{\left \{\begin{array}{l}
{\scriptstyle r = m\kappa  + n\kappa'} \\
 {\scriptstyle s= m\iota' + n\iota}\\
\end{array}\right.}.
\label{44}
\end{equation}
Both {\sl Ansatz} have as rationale the strategy of placing
the $N\equiv g_L/u$ electrons on a regular lattice. Moreover a peaked
distribution of the electron density around the sites diminishes
the energy coming from the Coulomb repulsion.
\par
Before examining the numerical consequences of the choices made in
eqs. (\ref{43}) and (\ref{44}), it is interesting to look closer
at the states occupied by the single electrons. In the state
$|\Omega\rangle$ the electron with $\mu\nu$ quantum numbers is
in a state given by (see eq. (\ref{38}))
\begin{equation}
\psi^{\mu\nu} = u^{-{1\over{2}}}
\sum_{\alpha\beta}^{{\rm fixed}~\mu\nu}
\phi^{\alpha\beta}
 = u^{-{1\over{2}}}
\sum_{\alpha\beta}^{{\rm fixed}~\mu\nu}\Vert\phi^{\alpha\beta}
\Vert\Phi^{\alpha\beta},
\label{45}
\end{equation}
while in the state $|{\tilde\Omega}\rangle$ (see eq. (\ref{41}))
is described by the wave function
\begin{equation}
u^{-{1\over{2}}}
\sum_{\alpha\beta}^{{\rm fixed}~\mu\nu}
\Phi^{\alpha\beta}.
\label{46}
\end{equation}
Thus in the first case the probability amplitude for the state
$\Phi^{\alpha\beta}$ is proportional to $\Vert\phi^{\alpha\beta}\Vert$
and in the second is equal for all $\alpha \beta$ (fixed $\mu\nu$).
\par
The two {\sl Ansatz} given in eqs. (\ref{43}) and (\ref{44})
can be
tested in the self-consistent equation based on the Hartree-Fock
approximation (see eq. (158) of III). Fig. \ref{w5} shows that
the state  $|\Omega\rangle$ given in  eq. (\ref{43}) yields a better
trial state (to a good approximation, it is unchanged by the
self-consistent procedure).
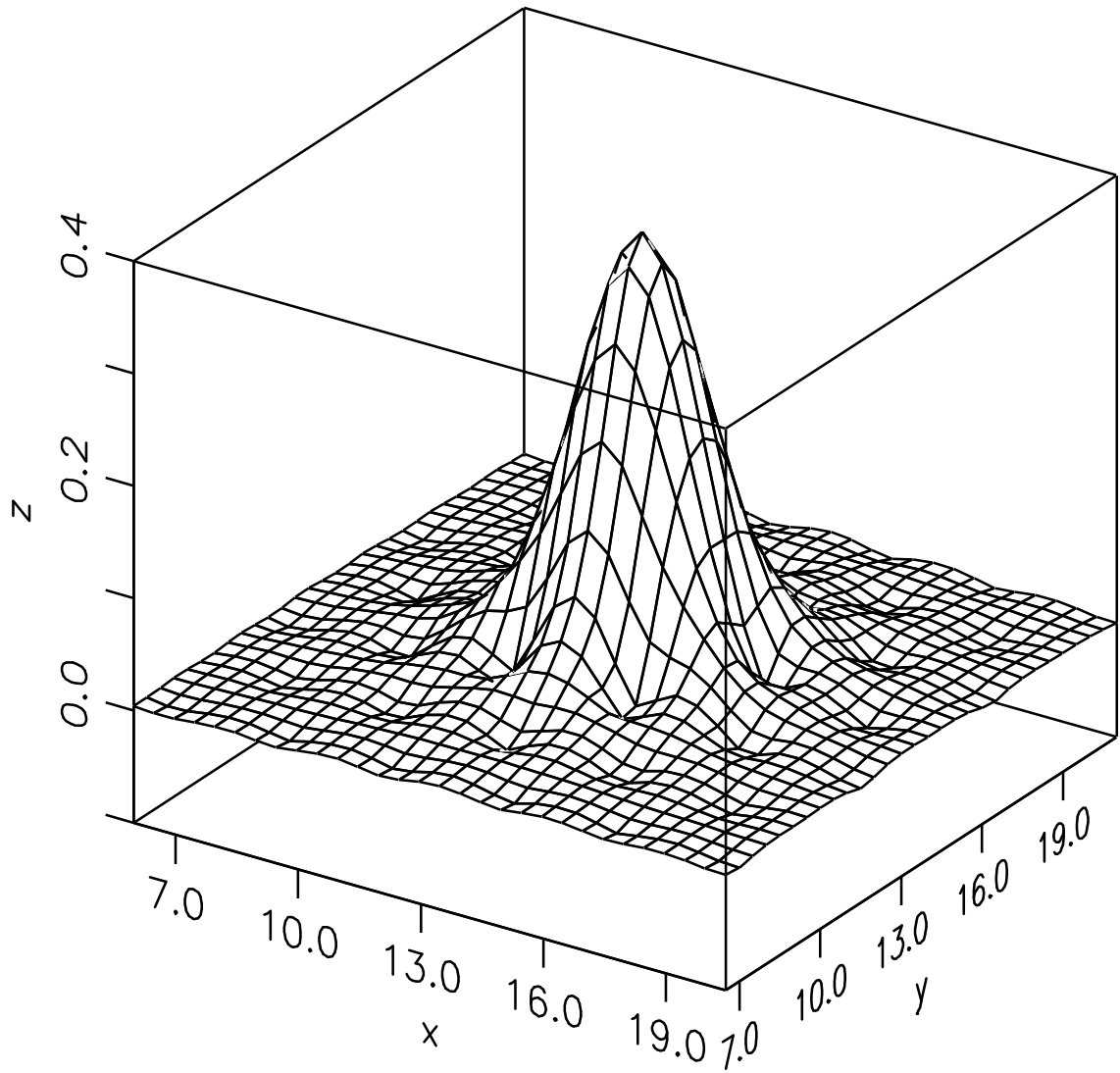
\begin{figure}
\caption{3D plot of the modulus of a Wannier function $\cal W$
given in eq. (35). The lattice is regular and triangular ($|f|=|g|
\sim 2.69$ in units of magnetic length). This picture is a zoom in
on a domain whose size appears in Figs. 2 and 3..
}
\label{w1}
\end{figure}
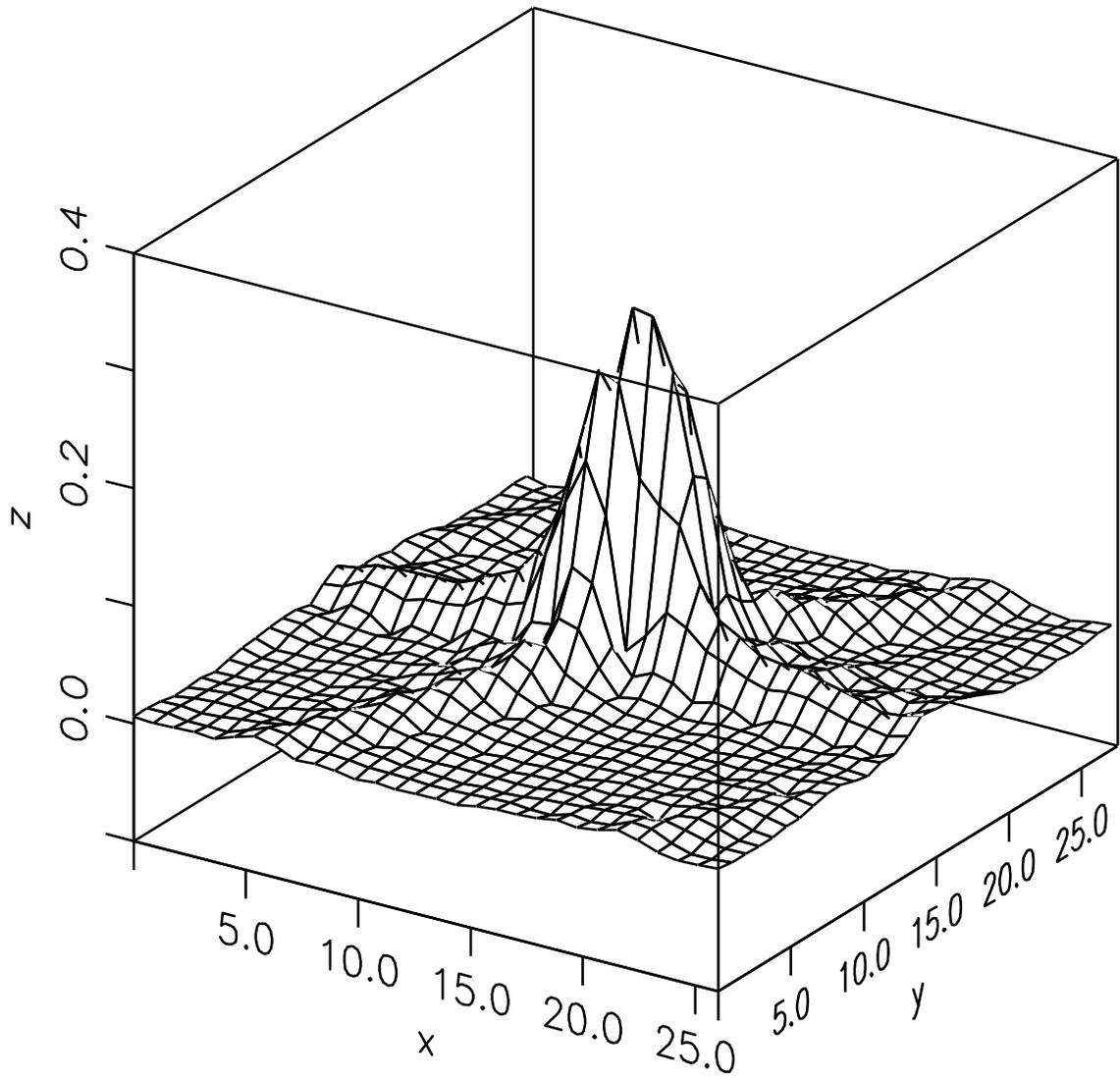
\begin{figure}
\caption{3D plot of the modulus of a Wannier function
$\hat{\cal W}$ given in eq. (36). See Fig. 1 for details.
}
\label{w2}
\end{figure}
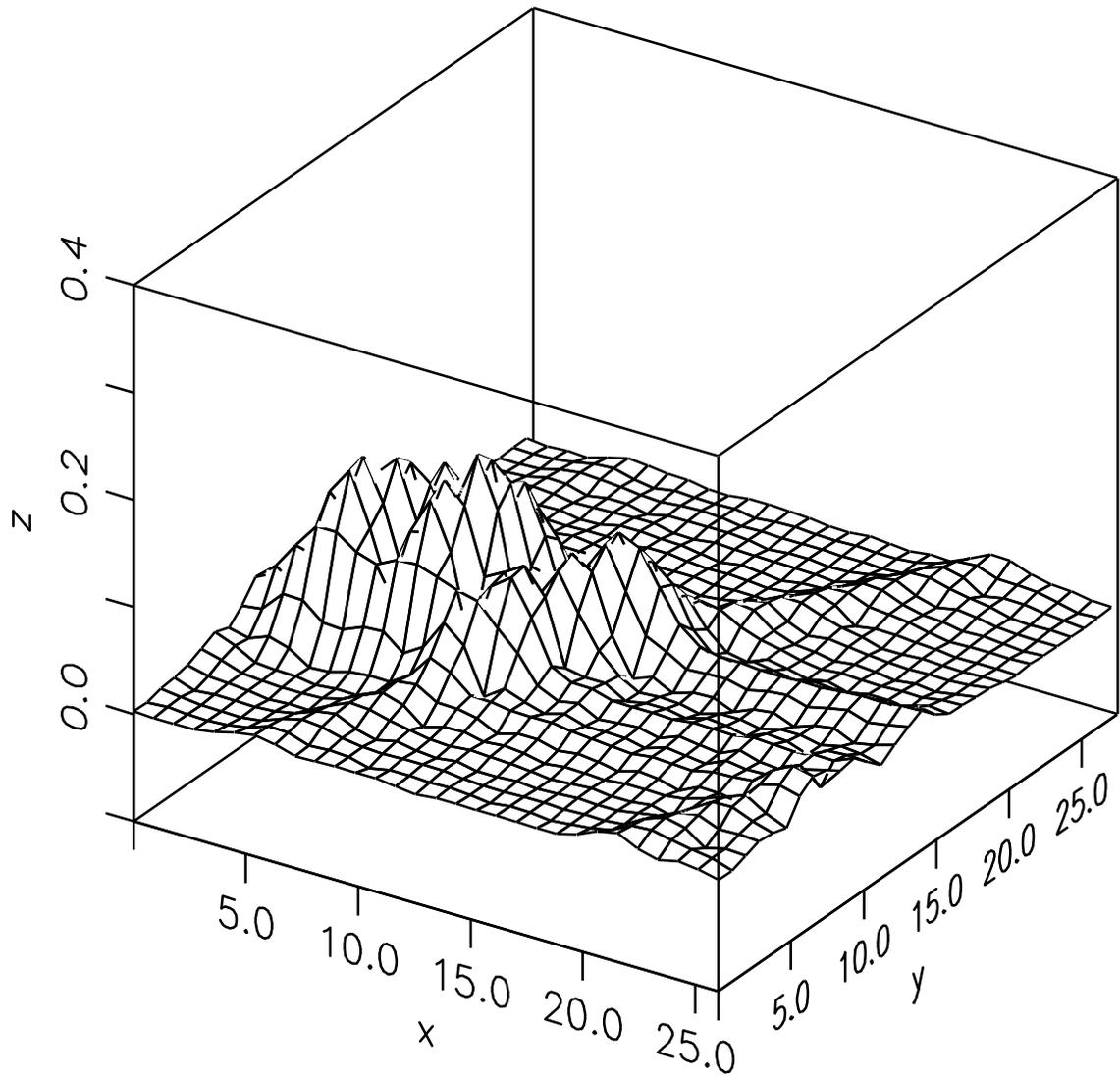
\begin{figure}
\caption{3D plot of the modulus of a Wannier function obtained
by using the representation of the MTG
given by the functions in eq. (37). See Fig. 1 for details.
}
\label{w3}
\end{figure}
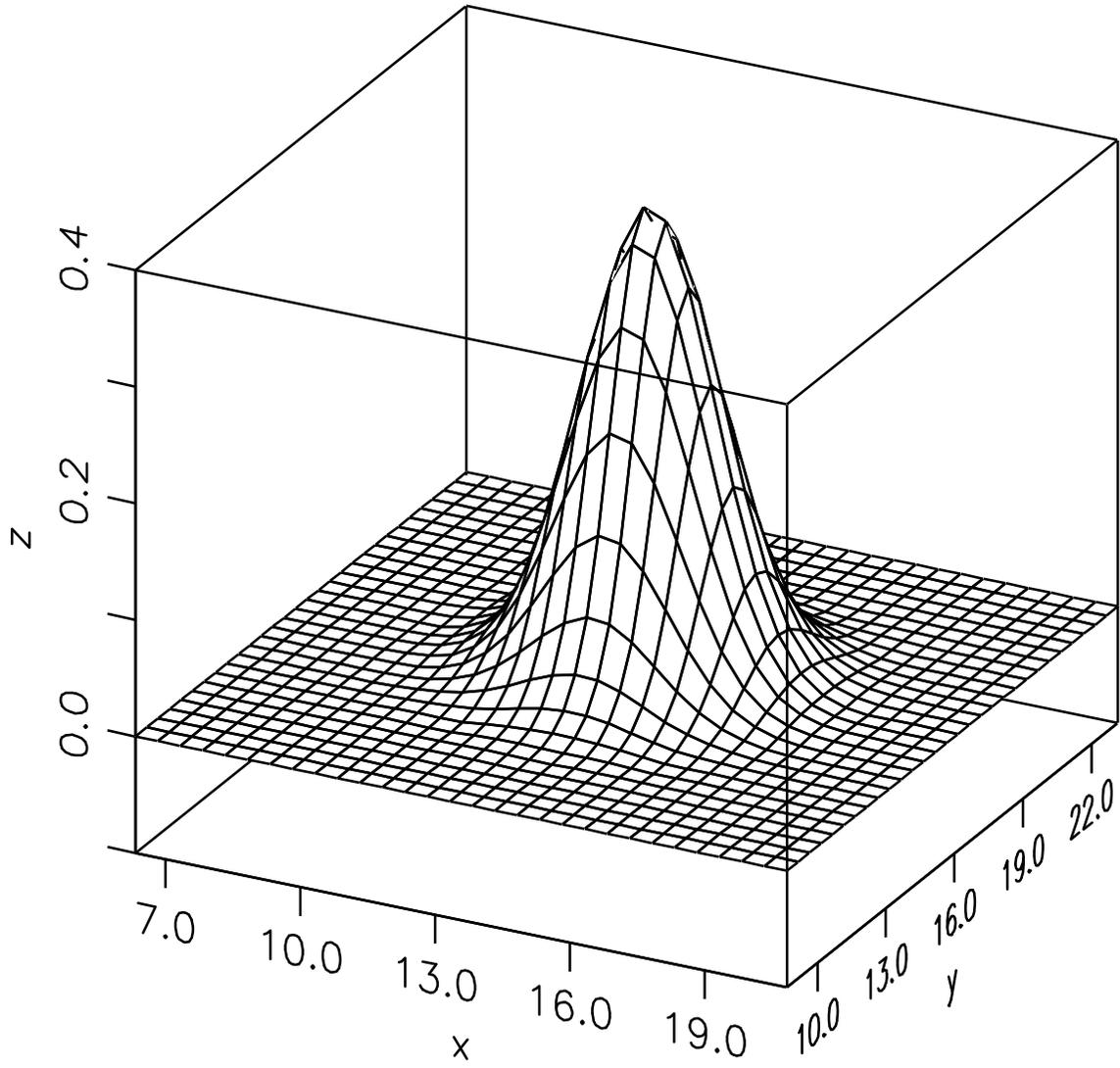
\begin{figure}
\caption{3D plot of the modulus of a Wannier function
${\cal W}^{(u)}$ given in eq. (40) ($(c,d)$ lattice, regular
and triangular. $u=3$ and therefore $|c|=|d|\sim 4.66$).
This picture is a zoom in on the domain of definition, whose
size appears in Figs. 2. and 3.}
\label{w4}
\end{figure}
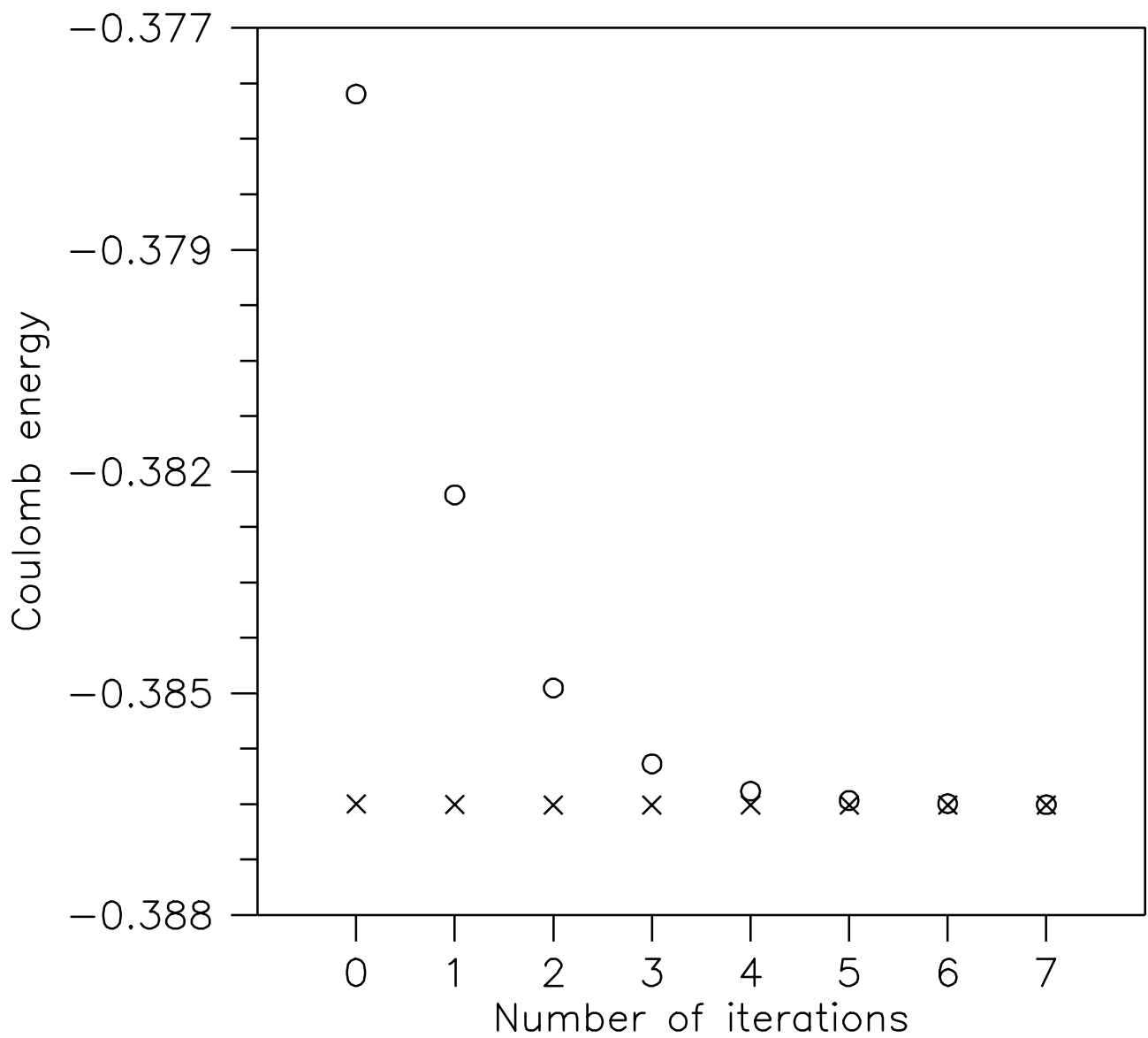
\begin{figure}
\caption{Convergence patterns in the iteration procedure based
on the self-consistent equation in the Hartree-Fock approximation.
"Times" refer to the {\sl Ansatz} based on the functions
 ${\cal W}^{(u)}$ (eq. (47))
and "circles" to the
{\sl Ansatz} based on the functions ${\cal W}$ (eq. (48)).
Filling factor is $1/3$.}
\label{w5}
\end{figure}
\end{document}